\documentclass[journal]{IEEEtran}
\usepackage{cite}
\usepackage{amsmath}
\usepackage{mathtools}
\usepackage{lipsum}
\usepackage{algorithmic}
\usepackage{array}
\usepackage{stfloats}
\usepackage{url}
\usepackage{xcolor}
\begin{document}

\title{Complex phase masks for fabricating OH-suppression filters for astronomy}

\author{Aashia Rahman, Kalaga Madhav, and Martin M. Roth%
\thanks{Aashia Rahman is with the innoFSPEC, Leibniz Institute for Astrophysics Potsdam (AIP), An der Sternwarte, Potsdam 14482, Germany e-mail: (arahman@aip.de).}% <-this % stops a space
\thanks{Kalaga Madhav is with the innoFSPEC, Leibniz Institute for Astrophysics Potsdam (AIP), An der Sternwarte, Potsdam 14482, Germany e-mail: (kmadhav@aip.de).}% <-this % stops a space
\thanks{Martin M Roth is with the innoFSPEC, Leibniz Institute for Astrophysics Potsdam (AIP), An der Sternwarte, Potsdam 14482, Germany e-mail: (mmroth@aip.de).}}% <-this % stops a space

\maketitle

\begin{abstract}
\ The design of a complex phase mask (CPM) for inscribing aperiodic filters in fibers is presented. The complex structure of the mask surface relief consists of discrete aperiodic phase-steps at periodic intervals. We show that the diffraction of the inscribing laser beam from the phase-step locations of the CPM produces periodically located half phase-steps along the fiber. The accumulated phase, along with controlled index modulation, generates the desired aperiodic reflection spectrum. Compared to a complex ``running-light" interferometry based inscription method, CPM offers the well known convenience and reproducibility of the standard phase mask inscription technique. The complexity of an aperiodic grating is encoded into the structure of the CPM. Complex filters fabricated with CPM can be used for suppressing the near infrared (NIR) OH-emission lines generated in the upper atmosphere, improving the performance of ground based telescopes.
\end{abstract}

\IEEEpeerreviewmaketitle

\section{Introduction}

\IEEEPARstart{G}{round}-based astronomical observations are adversely affected by various properties of the atmosphere, such as molecular absorption, extinction from dusts and aerosols, turbulence-induced wavefront aberrations, but also the overwhelming presence of sky background emission spanning 0.9 to 2.5$\mu$m, produced by hydroxyl (OH) radicals at an altitude of 90km \cite{Meinel}. Such sky emissions can be 1000x brighter than the science light from a distant object, e.g. a high redshift galaxy. While the interline continuum between the OH emission lines is intrinsically faint, the sensitivity of low to medium resolution NIR spectrographs between the lines is seriously degraded by the scattered-light induced broad wings of OH emission lines \cite{Maihara}. This is a very disturbing limitation for a new generation of extremely large telescopes, e.g. the ELT \cite{Ramsay}. As astronomy is photon-starved, with suppression of $>$25dB and interline transmission losses $<$0.5dB, a high order filter with negligible side-lobes is required.

Different methods like high dispersion masking, Rugate filters and holographics filters have been employed for OH-suppression; however, these methods inherently suffer from scattering properties due to dispersive optics and as a result, these methods were discovered to be inadequate when it comes to a complete suppression of the OH-lines \cite{Ellis}.

An extraordinary approach to suppress OH-lines in the H-band centered at 1.65$\mu$m, was first introduced by Bland-Hawthorn \emph{et al.} \cite{JBland} using fiber Bragg gratings (FBGs). FBG based filters bring in immense promise for OH-suppression as they are capable of filtering the OH-lines before the light enters a fiber-fed spectrograph \cite{ Ellis2, Trinh1, Ellis3}, hence prohibiting any contribution of scattered light in the spectrum.

Although a challenging task to realize, complex gratings can be fabricated using ultraviolet (UV) or femtosecond lasers by dephasing partial gratings \cite{Buryak} in optical fibers/waveguides \cite{Gbadebo, Zhu}. Point-by-point \cite{PbP}, line-by-line \cite{LbL} and  acousto - optic modulators based UV Talbot interferometry \cite{Gbadebo} have been explored to fabricate complex gratings. In these methods precise control on phase and intensity of the inscribing beam is required.

In what follows, we will advocate to adapt a popular method for inscribing gratings in optical fibers that uses off-the-shelf phase masks (PM) which are known for their high reproducibility. In ground-based observatories the light from the telescope is coupled into a multimode fiber. Using a photonic lantern \cite{Trinh1} the light is scrambled into multiple single mode fibers containing identical complex filters. Fabricating  identical filters on multiple fibers is challenging in interferometric inscription methods, where linear translation (x, y, z, pitch, roll, yaw) accuracy over a large length is in the order of few microns, even for air-bearing translation stages. By transferring half the complexity, in our case, the complex phase of the complex grating, onto a PM through a one-time manufacturing process, a CPM can be fabricated and can be used off-the-shelf in any standard PM fabrication setup, without requiring any high precision linear translation. In this paper, we introduce for the first time the design of such phase masks that can be used to fabricate OH-suppression filters in single mode optical fibers for ground based observatories. 

\section{Complex Mask Design}
\subsection{Synthesis of complex gratings}
\begin{figure}[htbp]
\centering
\includegraphics[width=90mm]{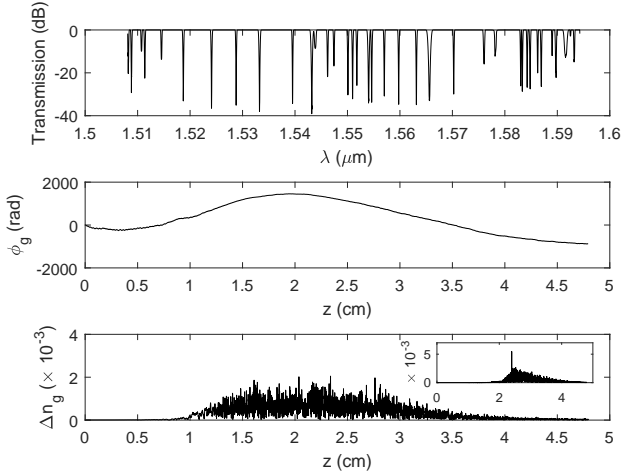}
\caption{Filter spectra consisting of 37 filter lines in H-band, and the phase $\phi_g$ and index profile $\Delta n_g$ of the complex grating. Inset (bottom) shows $\Delta n_g$ without dephasing.}
\label{fig:filter_n_p}
\end{figure}

The multichannel filter can be defined as \cite{Cao},

\begin{multline}\label{eq:11}
|r\big(\lambda\big)| =\sqrt{R_i}\sum_{i=1}^{N}{\exp \Bigg[-\Bigg(\frac{\lambda-\lambda_i}{\Delta\lambda_i}\Bigg)^{n}\Bigg]}\\
\times
 \exp \Bigg[i2\pi n_{e\!f\!f}\Bigg(\frac{1}{\lambda}-\frac{1}{\lambda_i}\Bigg)g_i\Bigg]
\end{multline}
where, $R_i$ is the desired reflectivity of $i^{th}$ channel, $\lambda_i$ is the central wavelength of $i^{th}$ channel, $N$ is the number of channels, $n$ defines the shape of the channel, $n_{e\!f\!f}$ is the effective refractive index, $\lambda_0$ is the seed grating, which also defines the CPM's seed pitch $\Lambda_m$, $\Delta\lambda_i$ is the full-width at half maximum (FWHM) of $i^{th}$ channel, and $g_i$ is the $i^{th}$ channel's group delay. $g_i$ is used to 'dephase' individual channels. We select $N$=37 emission lines in the H-band from 1508nm to 1593nm (bandwidth $F$=85nm), $\lambda_0$=1550nm and using \eqref{eq:11} we obtain the filter spectrum shown in Fig.\ref{fig:filter_n_p} (top).

The complex phase $\phi_g$ and index modulation $\Delta n_g$, given in Fig.\ref{fig:filter_n_p} (middle and bottom), of the physical grating in the fiber is derived from the desired reflection spectrum $|r(\lambda)|$ by using the Layer peeling (LP) method \cite{Skaar, Skaar1}. By dephasing individual channels we re-position the individual physical gratings along the fiber, essentially 'spreading' the complex grating along the full length $L$ of the grating. Without channel dephasing, all the gratings overlap at the same location in the fiber, increasing $\Delta n_g$ to values that cannot be achieved. Fig.\ref{fig:filter_n_p} (bottom) shows a comparison of $\Delta n_g$ with and without (inset) dephasing. $g_i$ is optimized using simulated annealing algorithm to further reduce the maximum index modulation in the fiber. The optimized $\Delta n_g$ and $\phi_g$ can now be used to design CPM.

\subsection{Design considerations}

$\Delta n_g$ can be inscribed by using an amplitude mask \cite{Mihailov}, acousto/electro-optic amplitude modulator, or by directly controlling the exposure time during the grating inscription. $\phi_g$ can be realized in two ways: (I) by sampling the grating using established sampling methods \cite{Wang, Qiang} or (II) as a non-linear chirp resulting from partially overlapping dephased individual gratings \cite{Buryak, Dai}.  Since $\Delta n_g$ can be implemented relatively easily compared to $\phi_g$, here we encode $\phi_g$ into the phase mask using method I.

\begin{figure}[htbp]
\centering
\includegraphics[width=\linewidth]{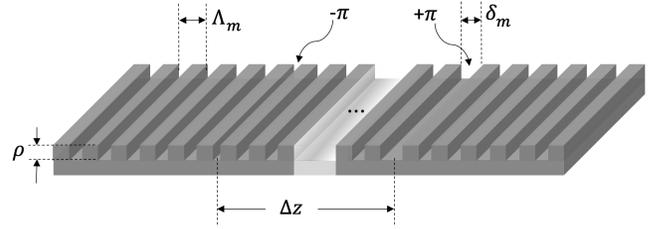}
\caption{Representative 3D model of a section of the CPM showing two $\Lambda_m/4$ shifted grooves corresponding to ($-\pi$, $+\pi$) phase, separated by $\Delta z$. $\rho=\lambda_{uv}/2(n_{uv}-1)$ is the groove depth of the phase mask, defined by the wavelength $\lambda_{uv}$ of the laser used for fabrication, and the refractive index $n_{uv}$ of the mask material.}
\label{fig:apm}
\end{figure}

\begin{figure}[htbp]
\centering
\includegraphics[width=50mm]{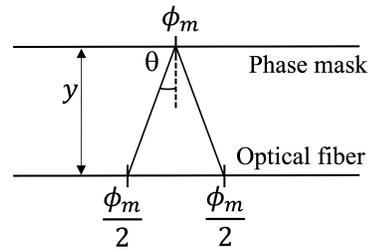}
\caption{Schematic showing the propagation of phase mask phase to fiber grating phase.}
\label{fig:mask_fiber1}
\end{figure}

The discrete step or the groove width of the CPM (Fig.\ref{fig:apm}) $\delta_m$ is given as,
\begin{equation}
\delta_m=\frac{\Lambda_{m}}{4\pi}\Big(2\pi+\phi_g\Big)
\label{eq:maskgap}
\end{equation}
We see that for $\delta_m$=$\frac{\Lambda_m}{4}$ (or $\frac{3\Lambda_m}{4}$), we get $\phi_g$=$-\pi$ (or $+\pi$) for a $\pi$-phase shifted mask. The discrete steps $\delta_m$ are recorded into the surface relief of the phase mask (Fig.\ref{fig:apm}) at intervals of $\Delta z=\frac{\lambda_0^2}{2n_{e\!f\!f}F}$ defined by the bandwidth $F$ of the total filter \cite{JBland3}. $\phi_g\in[-\pi,+\pi]\implies \delta_m\in[\frac{\Lambda_m}{4},\frac{3\Lambda_m}{4}]$. For the chosen seed grating    $\lambda_0\!=$1550nm, $\Lambda_m\!\approx$1064nm. In order to cover $-\pi$ to $+\pi$, the groove width $\delta_m$ in the CPM will range in between 266nm to 798nm. It is also important to note that the diffraction of orders of the mask at the location of the step $\delta_m$ splits the grating phase into two half-phases \cite{Sheng1}. Fig.\ref{fig:mask_fiber1} shows the projection of the mask phase to the grating plane; where, if $\phi_m$ is the phase shift in the mask, then in the core of the fiber two half-phase shifts $\frac{\phi_m}{2}$ are inscribed, separated by $2y\tan\theta$, where $y$ is the distance between the mask and the fiber core, and $\theta$ is the angle of diffraction of $\pm 1$ order. The discrete half-phases accumulate along the length of the fiber as light propagates through the grating, resulting in a desired complex phase $\phi_g$ of the filter.

\subsection{Fabrication constraints}
\begin{figure}[htbp]
\centering
\includegraphics[width=90mm]{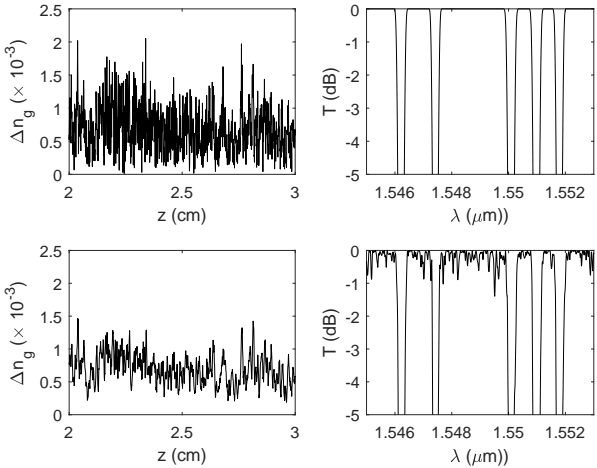}
\caption{Reducing the modulation index of $\Delta n_g$ results in a noisy filter spectrum even if the phase $\phi_g$ is completely fabricated into the complex filter. Top: preserving $\Delta n_g$. Bottom: Reduced modulation index. The continuum between 1545nm and 1553nm shows noise floor $>0.5$dB.}
\label{fig:filtered_n_p}
\end{figure}

The influence of $\Delta n_g$ and $\phi_g$ on the filter properties is numerically modelled to understand the fabrication constraints of using CPM in the standard phase mask inscription setup. In the first case we preserve $\phi_g$, and vary $\Delta n_g$ by reducing the modulation index. We see that if the details in $\Delta n_g$ are not preserved, even if the phase $\phi_g$ is inscribed without errors, the resulting filter spectra will be noisy (Fig.\ref{fig:filtered_n_p}), and the interline continuum, which contains the faint optical science light from the telescope is completely overwhelmed. 

Phase masks can be fabricated using e-beam lithography (EBL) and reactive ion etching (RIE) process. The resolution of the mask etching process, $\delta_e$, will result in quantization of $\delta_m$ over the range 266nm to 798nm. This in turn quantizes $\phi_g$. But with current state-of-the art RIE, the stitching errors in mask fabrication are in sub-nanometers, and have negligible influence on the fabricated filter spectra. However, to design filters with large bandwidth, we will require longer phase masks and minimising stitching errors in mask fabrication will be important. The need for longer phase masks can be circumvented by dividing a long-length phase mask into smaller phase masks each covering smaller bandwidths.

Inscribing $\phi_g$ and $\Delta n_g$ of the complex grating into the optical fiber requires a highly complex 'running light' Talbot-interferometer setup with sensitive optical alignment, inscription intensity control, and moving fiber mounts, where the velocity of translation is synchronized to the relative phases. In comparison, with the CPM we designed, where $\phi_g$ is pre-encoded through highly precise RIE processes, by controlling only the exposure time in a standard phase mask based fabrication setup, $\Delta n_g$ is easily inscribed into the fiber.

\section{Conclusion}
As OH-suppression becomes increasingly important for ground-based telescopes, the challenge in fabricating reproducible notch filters still remains, as the existing methods are experimentally tedious and cumbersome. We presented the design of a complex phase mask that reduces one of the key complexities, by pre-encoding the complex phase into the phase mask for inscribing complex multi-notch FBG filters. The key outcome of this work is substantially reducing the complexities in the inscription. Amplitude modulation along with CPM not only simplifies the process of inscription but also holds a promise for reproducible complex gratings fabrication which is essential for fiber-fed spectrographs for future extremely large ground-based telescopes.

\section*{Acknowledgment}
This work is supported by the BMBF project “Meta-ZiK\\ Astrooptics” (grant no. 03Z22A511). 

\ifCLASSOPTIONcaptionsoff
  \newpage
\fi

\bibliographystyle{IEEEtran}

 \newpage
\begin{IEEEbiographynophoto}{Aashia Rahman}
received her PhD in Instrumentation and Applied Physics from Indian Institute of Science, Bangalore in 2009, where she was also awarded the Dr. Srinivasa Rao Krishnamurthy Gold Medal. Following which she worked as Marie Sklodowska-Curie Fellow at the Electronic Structure and Laser (IESL), FORTH, Greece and, later she joined the Nanyang Technological University, Singapore as a postdoctoral research fellow. Afterwards, she served as an Assistant Professor in Indian Institute of Technology (IIT), Ropar, India. She later joined General Electric, India as a research analyst. Currently, she is a Senior Scientist at the Leibniz Institute for Astrophysics Potsdam (AIP), Germany.
\end{IEEEbiographynophoto}
\begin{IEEEbiographynophoto}{Kalaga Madhav}
received his PhD from Indian Institute of Science, Bangalore in 2007. After working as an assistant professor in Indian Institute of Technology (IIT), Ropar, India, as professor in VTU-CMRIT, India, and many years in industry, he joined the Leibniz Institute for Astrophysics Potsdam (AIP), Germany, where he is currently the group leader of Astrophotonics. The main research focus of his group is adaptive optics, sky suppression filters, compact spectrographs, beam combiners, and frequency combs for astronomy.
\end{IEEEbiographynophoto}
\begin{IEEEbiographynophoto}{Prof. Martin M Roth} is a professor at the University of Potsdam, Institute of Physics and Astronomy. He is a speaker for the innoFSPEC Potsdam Innovation Center and head of the innoFSPEC branch at the Leibniz Institute for Astrophysics Potsdam (AIP). He received his diploma degree in physics and his PhD in astrophysics from the Ludwig Maximilians-Universität Munich in 1986 and 1993, respectively. He is the author of more than 90 journal papers and has written two book chapters. His current research interests include resolved stellar populations in nearby galaxies, astrophotonics, and biomedical imaging.
\end{IEEEbiographynophoto}

\end{document}